
\magnification 1200
\line{\hfil RU-93-54}
\line{\hfil hep-th/9311165}

\vskip 0.4in
\centerline{\bf Local Magnetization in Critical Ising Model}
\centerline{\bf with Boundary Magnetic Field}
\vskip 0.6in
\centerline{R.Chatterjee \footnote\dag {email: robin@physics.rutgers.edu}
 and A.Zamolodchikov \footnote*{Supported in part by Department of Energy
under grant No. DE-FG0590ER40559}}

\centerline{Department of Physics and Astronomy}
\centerline{Rutgers University}
\centerline{P.O.Box 849, Piscataway, NJ 08855-0849}
\vskip 0.4in
\centerline{\bf Abstract}

We discribe a simple way to derive spin correlation functions in
2D Ising model at critical temperature but with nonzero magnetic field
at the boundary. Local magnetization (i.e. one-point function) is
computed explicitly for half-plane and disk geometries.

\vskip 0.6in

As is known, the continuous limit of 2D Ising model at critical temperature
$T = T_c$ is described by $c = 1/2$ conformal field theory which admits
a Lagrangian formulation as a theory of free massless Majorana Fermi
field $(\psi, \bar\psi)$ with the action

$${\cal A} = {{1\over {2\pi}}}\int d^2 z [\psi{\partial_{\bar z}}\psi
+ {\bar \psi}{\partial_z}{\bar \psi}]. \eqno(1)$$

Here $(z, \bar z) = (x + iy, x - iy)$ are complex coordinates of
euclidean plane and $d^2 z = dxdy$. The factor $-1/{2\pi}$ in this
action corresponds to standard normalization of the Fermi field, i.e.

$$\psi(z)\psi(z') = {1\over{z - z'}} + ...\,; \qquad {\bar\psi}(\bar
z){\bar\psi}(\bar z') = {1\over{\bar z - \bar z'}} + ...\, . \eqno(2)$$
The order parameter
field $\sigma(w, \bar w)$ is related to the above Fermi field
non-locally (it creates square-root branch cut for $\psi, \bar\psi$) so that
the the components $\psi(z), {\bar\psi}({\bar z})$ satisfy operator
product expansions

$$\psi(z)\sigma(w, \bar w) = (z - w)^{-{1\over 2}}\sum_{n=0}^{\infty}(z
- w)^{n}a_{-n}\sigma(w, \bar w); $$

$${\bar\psi}({\bar z})\sigma(w, \bar w) = ({\bar z} - {\bar w})^{-{1\over
2}}\sum_{n=0}^{\infty}({\bar z} - {\bar w})^{n}{\bar a}_{-n}\sigma(w,
\bar w), \eqno(3)$$
where $a_{-n}$ and ${\bar a}_{-n}$ are mode operators associated with
$\psi$ and $\bar \psi$. The field $a_{0}\sigma$ is known as ``disorder
parameter'' field $\mu$; in fact we have

$$a_{0}\sigma(w, \bar w) = {\omega\over{\sqrt 2}}\mu(w, \bar w);\qquad
{\bar a}_{0}\sigma(w, \bar w) = {{\bar \omega}\over{\sqrt 2}}\mu(w, \bar
w), \eqno(4)$$
where

$$\omega = e^{{{i\pi}\over 4}}; \qquad {\bar\omega} = e^{-{{i\pi}\over
4}}, \eqno(5)$$
and we have assumed the standard normalization of the fields
$\sigma$ and $\mu$, i.e.

$$\sigma (w, {\bar w}) \sigma (0, 0) = (w\bar w)^{-{1\over 8}}\times I +
...;\qquad \mu (w, {\bar w}) \mu (0, 0) = (w\bar w)^{-{1\over 8}}\times I +
..., \eqno(6)$$
where $I$ is the identity operator. In fact, the first two subleading terms in
(3) are expressed in terms of derivatives of $\mu$, namely

$$a_{-1}\sigma(w, \bar w) = {\omega\over{\sqrt 2}}(4{\partial_{w}}\mu (w,
\bar w));\quad a_{-2}\sigma(w, \bar w) = {\omega\over{\sqrt 2}}({8\over
3}{\partial_{w}^{2}}\mu (w, \bar w)). \eqno(7)$$
$${\bar a}_{-1}\sigma(w, \bar w) = {{\bar\omega}\over{\sqrt
2}}(4{\partial_{\bar w}}\mu (w, \bar w));\quad
{\bar a}_{-2}\sigma(w, \bar w) = {{\bar\omega}\over{\sqrt 2}}({8\over
3}{\partial_{\bar w}^{2}}\mu (w, \bar w)). \eqno(8)$$
These equations are easily obtained with the use of explicit expressions

$$T(z) = -{1\over 2}(\psi \partial_z \psi)(z); \qquad \bar T(\bar z) =
-{1\over 2}(\bar \psi \partial_{\bar z} \bar \psi)(\bar z) \eqno(9)$$
for the components $T = T_{zz}$ and $\bar T = T_{{\bar z}{\bar z}}$ of
stress-energy tensor in terms of Fermi fields $\psi, \bar \psi$.
Similar equations hold for the products $\psi \mu,\, \bar\psi \mu$;

$$\psi(z)\mu(w, \bar w) = {{\bar\omega}\over{\sqrt 2}}(z - w)^{-{1\over
2}}[\sigma (w, \bar w) + 4(z - w){\partial_w}\sigma (w, \bar w) + {8\over 3}(z
- w)^2 {\partial}_{w}^2 \sigma (w, \bar w) + ...];$$
$$\bar\psi(\bar z)\mu(w, \bar w) = {{\omega}\over{\sqrt 2}}(\bar
z - \bar w)^{-{1\over
2}}[\sigma (w, \bar w) + 4(\bar z - \bar w){\partial_{\bar w}}\sigma (w, \bar
w) + {8\over 3}(\bar z - \bar w)^2 {\partial}_{\bar w}^2 \sigma (w, \bar w)
+ ...]. \eqno(10)$$
This form of operator product expansions makes it possible to derive
linear differential equations which determine multipoint correlation
functions of the fields $\sigma$ and $\mu$(see e.g. [1]).

Conformal field theory of critical Ising model in the presence of
boundary (along with more general conformal field theories) is studied
in Refs[2-4]. It is shown in [3] that there are two essentially
different possibilities to choose ``critical'' boundary conditions, the
ones which preserve conformal invariance. If the boundary $\cal B$ is
given by a parametric curve

$${\cal B}:  \qquad z = Z(t); \quad \bar z = {\bar Z}(t), \eqno(11)$$
where $t$ is some real parameter along the boundary, these two
possibilities are

$$[e^{1\over 2}\psi - {\bar e}^{1\over 2}{\bar \psi}]_{\cal B} = 0
 \eqno(12)$$
and

$$[e^{1\over 2}\psi + {\bar e}^{1\over 2}{\bar \psi}]_{\cal B} = 0,
\eqno(13)$$
where

$$e(t)={d\over dt}Z(t), \qquad {\bar e(t)} = {d\over dt}{\bar Z}(t)
\eqno(14)$$
are the components of the vector $(e, \bar e)$ tangent to the boundary;
in what follows we assume that the parametrization (11) of the curve $\cal B$
is chosen in such a way that this  vector has  unit length, i.e.

$$e(t){\bar e}(t) = 1; \eqno(15)$$
we also assume the choice of orientation of this curve in which the
unit vector $(ie, -i{\bar e})$ normal to the boundary points inside the
domain $\cal D$ surrounded by $\cal B$; ${\cal B} = \partial {\cal D}$.
{}From a microscopic point of view, (12) and (13) correspond to ``free'' and
``fixed'' boundary conditions, respectively. Namely,
(12) appears when the microscopic boundary spins are not restricted,
while (13) corresponds to the situation when the boundary spins are fixed
to be all in the same position, $+1$ or $-1$ (so that (13) represents, in
fact, two boundary conditions, ``fixed to $+1$' and ``fixed to $-1$'').
In both cases, due to the conformal invariance, one can reduce the case
of a generic (one-component) curve $\cal B$ to the ``standard'' geometry where
$\cal D$ is the upper half-plane $y>0$. Then (12) and (13) simplify as

$$[\psi - \bar \psi]_{z=\bar z} = 0 \qquad for \quad ``free'' \quad case,
\eqno(12a)$$

$$[\psi + \bar \psi]_{z=\bar z} = 0 \qquad for \quad ``fixed'' \quad
case  \eqno(13a)$$
These equations show that the field $\bar \psi (\bar z)$ coincides (up
to a sign which depends on whether we choose ``free'' or ``fixed''
boundary condition) with
the analytic continuation of $\psi (z)$ to the lower half-plane.
Therefore it is still possible to use (7,8,10) to derive complete set of
linear differential equations which determine correlation functions of
the fields $\sigma$ and $\mu$ in the presence of boundary [2].

The above ``free'' and ``fixed'' boundary conditions are ``critical'',
i.e. they correspond to ``fixed points'' of the renormalization group
flow of ``boundary interactions'', and associated field theories with
these boundary conditions are conformally invariant. In this paper we
consider more
general boundary conditions which break conformal symmetry.
Specifically, we analyze the case when boundary spins interact with
constant external magnetic field $H_B$. Clearly, this boundary condition
``interpolates'' between the ``free'' case at $H_B = 0$ and ``fixed to
$\pm1$'' cases at $H_B \to \pm \infty$, and so it corresponds to
``boundary flow'' from ``free'' boundary condition down to ``fixed''
one[5]. This ``flow'' can be interpreted as ``free'' boundary condition
``perturbed'' by ``boundary spin operator'' $\sigma_B (t)$. The latter
was identified in [3,4] (see also [6]); it can be expressed in terms of
Fermi fields as

$$\sigma_B (t) = i\,a(t)[e^{1\over 2}\psi + {\bar e}^{1\over
2}{\bar \psi}](t), \eqno(16)$$
where $a(t)$ is auxiliary ``boundary'' Fermionic degree of freedom with
the two-point function $\langle a(t)a(t')\rangle_{free} =
{{1\over 2}}\, sign(t - t')\,$[6].
With this, the full action describing the continuous limit of $T=T_c$ Ising
model with boundary magnetic field takes the form

$${\cal A}_{h} = {1\over{2\pi}}\int_{\cal D}d^2 z [\psi{\partial_{\bar
z}}\psi + {\bar \psi}{\partial_z}{\bar \psi}] + \int_{\cal
B}[-{i\over{4\pi}}\psi{\bar \psi} + {1\over 2}a{\dot a}]dt + ih\int_{\cal
B}a(t)(e^{1\over 2}\psi + {\bar e}^{1\over 2}{\bar \psi})(t)dt, \eqno(17)$$
where $\dot a \equiv {d\over dt}a$. The first two terms here give the
action for Ising model field theory with `` free'' boundary condition
(10) and the last term is the ``perturbation'' describing the
interaction with the boundary field; it contains dimensional constant $h
\sim [length]^{-1}$ (appropriately rescaled external field $H_B$) and
breaks conformal symmetry. Note that the full action is quadratic, so it
is still a free field theory. The fact that the boundary magnetic field
does not destroy free-field structure of Ising model is very well known
(see e.g.[7]). Nevertheless, we consider it interesting to observe that
although conformal symmetry is broken the correlation functions
still satisfy linear differential equations.

The boundary condition for the fields $\psi, \bar \psi$ in the presence
of the boundary field $h$ is derived directly from (17),

$$({d\over dt} + i\lambda)\psi (t) = ({d\over dt} - i\lambda){\bar
\psi}(t), \eqno(18)$$

where

$$\psi(t) = e^{1\over 2}(t)\psi(Z(t)); \qquad {\bar \psi}(t) = {\bar
e}^{1\over 2}(t){\bar \psi}(Z(t)), \eqno(19)$$
and

$$\lambda = 4\pi h^2. \eqno(20)$$

We want to show that in the presence of boundary magnetic field $h$ the
correlation functions of $\sigma$ and $\mu$ still satisfy linear
differential equations. We will demonstrate this explicitely for
the one-point function $\langle \sigma (z, \bar z) \rangle_h$ (``local
magnetization'') in the simpest case of half-plane geometry. So, we
assume that the boundary $\cal B$ is the real axis $z = \bar z$ and
$\cal D$ is the upper half-plane $y > 0$. In this case the boundary
condition (19) reduces to

$$[(\partial_z + i\lambda)\psi(z) - (\partial_{\bar z} - i\lambda){\bar
\psi}(\bar z)]_{z=\bar z} = 0. \eqno(21)$$
This form of the boundary condition makes it explicit that the fields

$$\chi (z) = (\partial_z + i\lambda)\psi(z); \qquad {\bar \chi}(\bar z)
= (\partial_{\bar z} - i\lambda){\bar \psi}(\bar z), \eqno(22)$$
enjoy the desired property that $\bar \chi (\bar z)$ coincides with the
analytic continuation of $\chi (z)$ to the lower half-plane. So, for
instance, the correlation function $\langle \chi(z) \mu (w, \bar w)
\rangle$ is an analytic function of $z$ on the full $z$-plane with two
square-root branch points at $z = w$ and $z = \bar w$. Taking into
account (10) and the asymptotic behavior

$$\chi(z) \sim z^{-1} \quad as \quad z \to \infty \eqno(23)$$
one can write

$$\langle \chi(z) \mu (w, \bar w)\rangle_h (z - w)^{1\over 2}(z - \bar
w)^{1\over 2} = {{A(w, \bar w)}\over {z - w}} + {{{\bar A}(w, \bar
w)}\over {z - \bar w}} + B(w, \bar w). \eqno(24)$$
Now, one can use the operator product expansions

$$\chi(z)\mu(w, \bar w) = {{\bar\omega}\over{\sqrt 2}}(z - w)^{-{3\over
2}}(-{1\over 2}\sigma(w, \bar w) + (z - w)(2\partial_w +
i\lambda)\sigma(w, \bar w) + $$
$$+ (z - w)^{2}(4\partial_{w}^{2} +
4i\lambda\partial_{w})\sigma(w, \bar w) + ...); \eqno(25)$$
and

$${\bar\chi}(\bar z)\mu(w, \bar w) = {{\omega}\over{\sqrt 2}}(\bar z -
\bar w)^{-{3\over 2}}(-{1\over 2}\sigma(w, \bar w) + (\bar z - \bar
w)(2\partial_{\bar w} - i\lambda)\sigma(w, \bar w) + $$
$$+ (\bar z - \bar w)^{2}(4\partial_{\bar w}^{2} -
4i\lambda\partial_{\bar w})\sigma(w, \bar w) + ...), \eqno(26)$$
which follow from (10), to express the coefficients $A, \bar A$ and $B$
in (24) in terms of the function $\langle \sigma (w, \bar w) \rangle$
and its derivatives. For instance, expanding (24) in powers of $z - w$
with the use of (25) we get

$$A(w, \bar w) = {{\bar\omega}\over{\sqrt 2}}(w - \bar w)^{1\over 2}
(-{1\over 2})\langle \sigma (w, \bar w) \rangle_h; \eqno(27a)$$
$$B(w, \bar w) + {{\bar A(w, \bar w)}\over {w - \bar w}} =
{{\bar\omega}\over{\sqrt 2}}(w - \bar w)^{1\over 2}(2\partial_{w} +
i\lambda - {1\over 4}{1\over{w - \bar w}})\langle \sigma(w, \bar w)
\rangle_h; \eqno(27b)$$
$$-{{{\bar A}(w, \bar w)}\over{(w - \bar w)^2}}
={{\bar\omega}\over{\sqrt 2}}(w - \bar w)^{1\over 2}(4\partial_{w}^{2} +
(4i\lambda + {1\over {w - \bar w}})\partial_{w} + $$
$$+{1\over 2}{i\lambda\over {w - \bar w}} + {1\over 16}{1\over {(w -
\bar w)^2}})\langle \sigma(w, \bar w) \rangle_h. \eqno(27c)$$
On the other hand, in view of (21), the expansion of this function
in powers of $z - \bar w$ is controlled by the operator product
expansion (26). Therefore we have also

$${\bar A}(w, \bar w) = {{\omega}\over{\sqrt 2}}(\bar w - w)^{1\over 2}
(-{1\over 2})\langle \sigma (w, \bar w) \rangle_h; \eqno(28a)$$
$$B(w, \bar w) + {{A(w, \bar w)}\over {\bar w - w}} =
{{\omega}\over{\sqrt 2}}(\bar w - w)^{1\over 2}(2\partial_{\bar w} -
i\lambda - {1\over 4}{1\over{\bar w - w}})\langle \sigma(w, \bar w)
\rangle_h; \eqno(28b)$$
$$-{{{A}(w, \bar w)}\over{(\bar w - w)^2}}
={{\omega}\over{\sqrt 2}}(\bar w - w)^{1\over 2}(4\partial_{\bar w}^{2} +
(-{4i\lambda} + {1\over {\bar w - w}})\partial_{\bar w} - $$
$$-{1\over 2}{i\lambda\over {\bar w - w}} + {1\over 16}{1\over {(\bar w -
w)^2}})\langle \sigma(w, \bar w) \rangle_h. \eqno(28c)$$
Compatibility of (27a,b) and (28a,b) requires that

$$(\partial_w + \partial_{\bar w})\langle \sigma(w, \bar w) \rangle_h = 0.
\eqno(29)$$
This equation just expresses the translational symmetry of the system:
the one-point function $\langle \sigma(w, \bar w) \rangle_h$ does not
depend on $x = {1\over 2}(w + \bar w)$. With this, (28c) becomes
identical to (27c) and reduces to the ordinary second-order differential
equation

$$[4{{d^2}\over{dY^2}} + ({1\over Y} - 4){d\over {dY}} + (-{1\over Y} +
{9\over 16}{1\over Y^2})]{\bar \sigma}(Y) = 0 \eqno(30)$$
for the function

$${\bar \sigma}(2\lambda y) = \langle \sigma (iy, -iy) \rangle_h .
\eqno(31)$$
This can be reduced to degenerate hypergeometric equation and we find
that the only solution which does not grow exponentially as $Y \to
+\infty$ is

$${\bar \sigma}(Y) = C \times {Y^{3\over 8}}\Psi ({1/2}, 1,
Y), \eqno(32)$$
where

$$\Psi (a, c, x) = {1\over {\Gamma(a)}}\int_{0}^{\infty}e^{-xt}t^{a-1}(1
+ t)^{c-a-1}dt. \eqno(33)$$
The $Y$-independant constant $C$ can be fixed by comparing the $Y \to 0$
asymptotic of the solution (32)

$${\bar \sigma}(Y) = C \times {1\over{\sqrt{\pi}}}{Y^{3\over 8}}[-\log (Y) +
O(1)] \eqno(34)$$
with the first nontrivial perturbative contribution

$$\langle \sigma (iy, -iy) \rangle_h = h\int_{-L}^{+L}\langle
\sigma (iy, -iy) \sigma_B (x) \rangle_{free}dx + O(h^3) = $$
$$= 2^{5\over 4} h (2y)^{3\over 8}(\log(2L) - \log (2y)) + O(h^3);
\eqno(35)$$
where we introduced an infrared cut-off $L$ and assumed that the field
$\sigma$ is normalized as in (6). Finally,

$$\langle \sigma (iy, -iy) \rangle_h = {\lambda}^{1\over 2} 2^{1\over 4}
(2y)^{3\over 8}\Psi (1/2, 1, 2\lambda y). \eqno(36)$$
Note that the $Y \to \infty$ asymptotic

$$\langle \sigma (iy, -iy) \rangle_h \sim 2^{1\over 4}(2y)^{-{1\over 8}}
\eqno(37)$$
of this function coincides with $\langle \sigma (iy, -iy)
\rangle_{fixed}$(see [4]), as it should, since $h \to \infty$ limit
corresponds to the ``fixed to +1'' boundary condition.

The above analysis can be repeated in the case when $\cal D$ is a disk
of radius $R$, i.e. the boundary $\cal B$ is the circle

$$Z(t) = R \exp({{it}\over {2\pi R}}); \qquad  \bar Z (t) =
R \exp(-{{it}\over {2\pi R}}). \eqno(38)$$
Obviously, in this case the one-point function $\langle \sigma(z, \bar z)
\rangle_{h}$ depends only on the combination $z\bar z$, i.e.

$$\langle \sigma(z, \bar z)\rangle_{h} =  s({{z\bar z}/{R^2}}),
\eqno(39)$$
and the function $s(X)$ can be shown to satisfy

$$[4{d^{2}\over dX^2} + ({{4(1+g)}\over{X}} - {1\over{1-X}}){d\over dX}
+ ({9\over 16}{1\over{(1-X)^2}} + {{1-g}\over{2X(1-X)}})]s(X) = 0,
\eqno(40)$$
where $g = \lambda R$. This equation is solved in terms of
hypergeometric functions; the solution which is regular everywhere
inside the disk $\cal D$ is

$$\langle \sigma(z, \bar z)\rangle_{h} =g^{1\over
2}{{\Gamma(1/2+g)}\over{\Gamma(1+g)}}(R/4)^{-{1\over 8}}(1-X)^{3\over
8}F(1/2, 1/2 + g, 1 + g, X), \eqno(41)$$
where

$$X = {{z\bar z}\over {R^2}}, \qquad g = 4\pi h^2 R, \eqno(42)$$
and $F(a, b, c, X))$ is the hypergeometric function. Again, we used the
first nontrivial order of perturbation theory in $h$ to fix the overall
normalization factor in (41). In the limit when $R$ goes to infinity
while the distance from $(z, \bar z)$ to $\cal B$ is kept finite (i.e.
$X \to 1$ with $Y = \lambda R (1-X)$ fixed), (41) degenerates to (32).

Similar arguments can be used to derive differential equations for
multipoint functions. However, these equations are more complicated and
we do not describe them here.

\vskip 0.6in

\centerline{\bf Acknowledgement}
\vskip 0.2in
RC would like to thank V.Brazhnikov and Y.Pugay for helpful discussions.
\vfill\eject
\vskip 0.6in

\centerline{\bf References}
\vskip 0.2in
1. ``Conformal Invariance and Applications to Statistical Mechanics'',

C.Itzykson, H.Saleur, J.-B.Zuber, eds., World Scientific, 1988.

2.J.Cardy. ``Conformal Invariance and Surface Critical Behavior'', in
Ref.1.

3.J.Cardy. Nucl.Phys. B324 (1989) p.581.

4.J.Cardy, D.Lewellen. Phys.Lett. 259B (1991) p.274.

5.I.Affleck, A.Ludwig. ``Universal Non-Integer ``Groundstate

Degeneracy'' in Critical Quantum Systems'', Preprint UBCTP-91-007.

6.S.Ghoshal, A.Zamolodchikov. ``Boundary S-matrix and Boundary State in

Two-Dimensional Integrable Quantum Field Theory''. Preprint RU-93-20.

7.B.M.McCoy, T.T.Wu. ``The Two-Dimensional Ising Model'', Harvard

University Press, 1973.
\end